 \documentclass[preprint,12pt]{elsarticle}

\usepackage{graphics}

\usepackage{amssymb}

\usepackage{lineno}

\usepackage{upgreek}

\journal{Physics Letter B}

\def\ifm#1{\relax\ifmmode#1\else$#1$\fi}

\def\fig#1{Figure~\ref{#1}}
\def\tab#1{Table~\ref{#1}}
\def\equ#1{eq.~(\ref{#1})}

\def\Ref#1{Ref.~\cite{#1}}

\def\etal{{\it et al.}}

\newcommand{\affuni}[2]{Dipartimento di Fisica dell'Universit\`a #1, #2, Italy.}
\newcommand{\affinfn}[2]{INFN Sezione di #1, #2, Italy.}

\def\f{\ifm{\upphi}}    
\def\ff{\f--factory}
\renewcommand{\to}{\ensuremath{\rightarrow}}
\def\dafne{DA$\Upphi$NE}
\def\ep{\ifm{\mathrm{e}^+}}
\def\el{\ifm{\mathrm{e}^-}}
\def\ks{\ifm{\mathrm{K}_{\mathrm{S}}}}
\def\kl{\ifm{\mathrm{K}_{\mathrm{L}}}}
\def\ksl{\ifm{\mathrm{K}_{\mathrm{S,L}}}}
\def\ko{\ifm{\mathrm{K}^0}}
\def\ok{\ifm{{\bar{\mathrm{K}}}^0}}

\def\pion{\ifm{\uppi}}
\def\pip{\ifm{\uppi^+}}
\def\pim{\ifm{\uppi^-}}

\def\JPC{\ifm{J^{PC}}}
\def\T{\ifm{\mathrm{T}}}
\def\C{\ifm{\mathrm{C}}}
\def\P{\ifm{\mathrm{P}}}
\def\CP{\ifm{\mathrm{CP}}}
\def\CPT{\ifm{\mathrm{CPT}}}

\def\ketre{\ifm{\mathrm{K}_{e3}}}
\def\muon{\ifm{\upmu}}

\def\mum{\ifm{\upmu^-}}
\def\neutrino{\ifm{\upnu}}
\def\ab{\ifm{\sim}}
\newcommand{\ket}[1]{\ifm{|#1\rangle}}
\newcommand{\braket}[2]{\ifm{\langle #1|#2 \rangle}}
\def\order#1,{\ifm{\mathcal{O}(10^{#1})}}
\def\x{\ifm{\times}}
\def\dd{\ifm{\mathrm{d}}}
\newcommand{\sinp}[2]{\ifm{\sin^{\!#1}\!#2}}

\def\MeV{\ifm{\mathrm{MeV}}}
\def\GeV{\ifm{\mathrm{GeV}}}
\def\mum{\ifm{\upmu\mathrm{m}}}
\def\mm{\ifm{\mathrm{mm}}}

\def\invfb{\ifm{\mathrm{fb}^{-1}}}

\def\Dt{\ifm{\Delta \tau}}
\def\Ts{\ifm{{t_\mathrm{s}}}}

\def\Damu{\ifm{\Delta a_\mu}}
\def\Dax{\ifm{\Delta a_\mathrm{X}}}
\def\Day{\ifm{\Delta a_\mathrm{Y}}}
\def\Daz{\ifm{\Delta a_\mathrm{Z}}}
\def\Dao{\ifm{\Delta a_0}}
\def\puno{\ifm{\vec{p}_1}}
\def\pdue{\ifm{\vec{p}_2}}
\def\pfi{\ifm{{\vec{p}}_\upphi}}
\def\Ipm{\ifm{I_{\pm}(\Dt)}}
\def\deltak{\ifm{\delta_{\mathrm{K}}}}

\begin{document}


\begin{frontmatter}




\title{Test of \CPT\ and Lorentz symmetry  
  in entangled neutral kaons with the KLOE experiment}


\author{The KLOE-2 Collaboration\\}
\author[Frascati]{D.~Babusci}
\author[Cracow]{I.~Balwierz-Pytko}
\author[Frascati]{G.~Bencivenni}
\author[Frascati]{C.~Bloise}
\author[Frascati]{F.~Bossi}
\author[INFNRoma3]{P.~Branchini}
\author[Roma3,INFNRoma3]{A.~Budano}
\author[Uppsala]{L.~Caldeira~Balkest\aa hl}
\author[Frascati]{G.~Capon}
\author[Roma3,INFNRoma3]{F.~Ceradini}
\author[Frascati]{P.~Ciambrone}
\author[Messina,INFNCatania]{F.~Curciarello}
\author[Cracow]{E.~Czerwi\'nski}
\author[Frascati]{E.~Dan\`e}
\author[Messina,INFNCatania]{V.~De~Leo}
\author[Frascati]{E.~De~Lucia}
\author[INFNBari]{G.~De~Robertis}
\author[Roma1,INFNRoma1]{A.~De~Santis\corref{cor1}}
\ead{antonio.desantis@roma1.infn.it}
\ead[url]{http://www.lnf.infn.it/\~adesanti}
\cortext[cor1]{Corresponding author}
\author[Frascati]{P.~De~Simone}
\author[Roma3,INFNRoma3]{A.~Di~Cicco}
\author[Roma1,INFNRoma1]{A.~Di~Domenico}
\author[Napoli,INFNNapoli]{C.~Di~Donato}
\author[INFNRoma2]{R.~Di~Salvo}
\author[Frascati]{D.~Domenici}
\author[Bari,INFNBari]{O.~Erriquez}
\author[Bari,INFNBari]{G.~Fanizzi}
\author[Roma2,INFNRoma2]{A.~Fantini}
\author[Frascati]{G.~Felici}
\author[Roma1,INFNRoma1]{S.~Fiore}
\author[Roma1,INFNRoma1]{P.~Franzini}
\author[Cracow]{A.~Gajos}
\author[Roma1,INFNRoma1]{P.~Gauzzi}
\author[Messina,INFNCatania]{G.~Giardina}
\author[Frascati]{S.~Giovannella}
\author[INFNRoma3]{E.~Graziani}
\author[Frascati]{F.~Happacher}
\author[Uppsala]{L.~Heijkenskj\"old}
\author[Uppsala]{B.~H\"oistad}
\author[Uppsala]{M.~Jacewicz}
\author[Uppsala]{T.~Johansson}
\author[Cracow]{K.~Kacprzak}
\author[Cracow]{D.~Kami\'nska}
\author[Uppsala]{A.~Kupsc}
\author[Frascati,StonyBrook]{J.~Lee-Franzini}
\author[INFNBari]{F.~Loddo}
\author[Roma3,INFNRoma3]{S.~Loffredo}
\author[Messina,INFNCatania,CentroCatania]{G.~Mandaglio}
\author[Moscow]{M.~Martemianov}
\author[Frascati,Marconi]{M.~Martini}
\author[Roma2,INFNRoma2]{M.~Mascolo}
\author[Roma2,INFNRoma2]{R.~Messi}
\author[Frascati]{S.~Miscetti}
\author[Frascati]{G.~Morello}
\author[INFNRoma2]{D.~Moricciani}
\author[Cracow]{P.~Moskal}
\author[INFNRoma3,LIP]{F.~Nguyen}
\author[Frascati]{A.~Palladino}
\author[INFNRoma3]{A.~Passeri}
\author[Energetica,Frascati]{V.~Patera}
\author[Roma3,INFNRoma3]{I.~Prado~Longhi}
\author[INFNBari]{A.~Ranieri}
\author[Frascati]{P.~Santangelo}
\author[Frascati]{I.~Sarra}
\author[Calabria,INFNCalabria]{M.~Schioppa}
\author[Frascati]{B.~Sciascia}
\author[Cracow]{M.~Silarski}
\author[Roma3,INFNRoma3]{C.~Taccini}
\author[INFNRoma3]{L.~Tortora}
\author[Frascati]{G.~Venanzoni}
\author[Warsaw]{W.~Wi\'slicki}
\author[Uppsala]{M.~Wolke}
\author[Cracow]{J.~Zdebik}
\address[Bari]{\affuni{di Bari}{Bari}}
\address[INFNBari]{\affinfn{Bari}{Bari}}
\address[CentroCatania]{Centro Siciliano di Fisica Nucleare e Struttura della Materia, Catania, Italy.}
\address[INFNCatania]{\affinfn{Catania}{Catania}}
\address[Calabria]{\affuni{della Calabria}{Cosenza}}
\address[INFNCalabria]{INFN Gruppo collegato di Cosenza, Cosenza, Italy.}
\address[Cracow]{Institute of Physics, Jagiellonian University, Cracow, Poland.}
\address[Frascati]{Laboratori Nazionali di Frascati dell'INFN, Frascati, Italy.}
\address[Messina]{Dipartimento di Fisica e Scienze della Terra 
  dell'Universit\`a di Messina, Messina, Italy.}
\address[Moscow]{Institute for Theoretical and Experimental Physics (ITEP), 
  Moscow, Russia.}
\address[Napoli]{\affuni{``Federico II''}{Napoli}}
\address[INFNNapoli]{\affinfn{Napoli}{Napoli}}
\address[Energetica]{Dipartimento di Scienze di Base ed Applicate per 
  l'Ingegneria dell'Universit\`a ``Sapienza'', Roma, Italy.}
\address[Marconi]{Dipartimento di Scienze e Tecnologie applicate, 
  Universit\`a ``Guglielmo Marconi", Roma, Italy.}
\address[Roma1]{\affuni{``Sapienza''}{Roma}}
\address[INFNRoma1]{\affinfn{Roma}{Roma}}
\address[Roma2]{\affuni{``Tor Vergata''}{Roma}}
\address[INFNRoma2]{\affinfn{Roma Tor Vergata}{Roma}}
\address[Roma3]{Dipartimento di Matematica e Fisica dell'Universit\`a 
  ``Roma Tre'', Roma, Italy.}
\address[INFNRoma3]{\affinfn{Roma Tre}{Roma}}
\address[StonyBrook]{Physics Department, State University of New 
York at Stony Brook, USA.}
\address[Uppsala]{Department of Physics and Astronomy, Uppsala University, 
  Uppsala, Sweden.}
\address[Warsaw]{National Centre for Nuclear Research, Warsaw, Poland.}
\address[LIP]{Present Address: Laborat\'orio de Instrumenta\c{c}\~{a}o e 
  F\'isica Experimental de Part\'iculas, Lisbon, Portugal.}

\begin{abstract}
Neutral kaon pairs produced in \f\ decays in anti-symmetric entangled 
state can be exploited to search for violation of \CPT\ symmetry and
Lorentz invariance. We present an analysis of the \CP-violating process
\f\to\ks\kl\to\pip\pim\pip\pim\ based on 1.7 \invfb\ of data collected 
by the KLOE experiment at the Frascati \ff\ \dafne.
The data are used to perform a measurement of the \CPT-violating 
parameters \Damu\ for neutral kaons in the context of the Standard Model
Extension framework. The parameters measured in the reference frame of 
the fixed stars are: \\
$\Dao = (-6.0 \pm 7.7_{stat} \pm 3.1_{syst})\times 10^{-18} \,\, \GeV$  \\
$\Dax = (\,\,\,0.9 \pm 1.5_{stat} \pm 0.6_{syst})\times 10^{-18} \,\, \GeV$  \\
$\Day = (-2.0 \pm 1.5_{stat} \pm 0.5_{syst})\times 10^{-18} \,\,\GeV$  \\
$\Daz = (\,\,\,3.1 \pm 1.7_{stat} \pm 0.5_{syst})\times 10^{-18} \,\,\GeV$  \\
These are presently the most precise measurements in the quark sector of 
the Standard Model Extension.
\end{abstract}


\end{frontmatter}









\section{Introduction} \label{sec:intro}

The \CPT\ theorem \cite{cpttheo} ensures invariance under the 
simultaneous transformation of charge conjugation (\C), parity (\P) and time 
reversal (\T), in the context of local quantum field theories with Lorentz 
invariance and Hermiticity.
On the other hand \CPT\ violation in any unitary, local, point-particle 
quantum field theory entails Lorentz invariance violation, as proved in a 
theorem by Greenberg~\cite{Greenberg:2002uu}.
Even though Lorentz symmetry has been verified in many experiments, as also 
\CPT\ invariance, it can be considered a 
sensitive probe for physics at the Planck scale, where natural mechanisms 
for such violations exist in quantum theories of \
gravity~\cite{gravity}.
\par
An attractive possibility to describe \CPT\ and Lorentz violation effects
in the low energy regime accessible to experiments consists of using an
effective field theory, independent of the underlying mechanism generating 
\CPT\ and Lorentz violation. This can be obtained by adding to the Standard 
Model action all possible scalar terms formed by contracting operators for 
simultaneous \CPT\ and Lorentz violation with coefficients that control the 
size of the effects. 
The resulting effective field theory, the Standard Model 
Extension (SME)\cite{Colladay:1996iz}, appears to be compatible with the 
basic tenets of quantum field theory and retains the properties of gauge
invariance and renormalizability. The SME is quite general and widely used in 
a broad class of experimental tests of \CPT\ and Lorentz symmetry in several 
physics domains ranging from atomic physics to particle physics and to 
cosmology~\cite{Kostelecky:2008ts}.
\par
In the present Letter we present a test of \CPT\ and Lorentz symmetry in the 
neutral kaon system within the SME framework. It has been performed by 
analyzing 
the \f\to\ks\kl\to\pip\pim\,\pip\pim\ events collected by the KLOE experiment 
at the \dafne\ \ff.
\par
\begin{figure}[htb]
  \begin{center}    
    \includegraphics[width=0.40\textwidth]{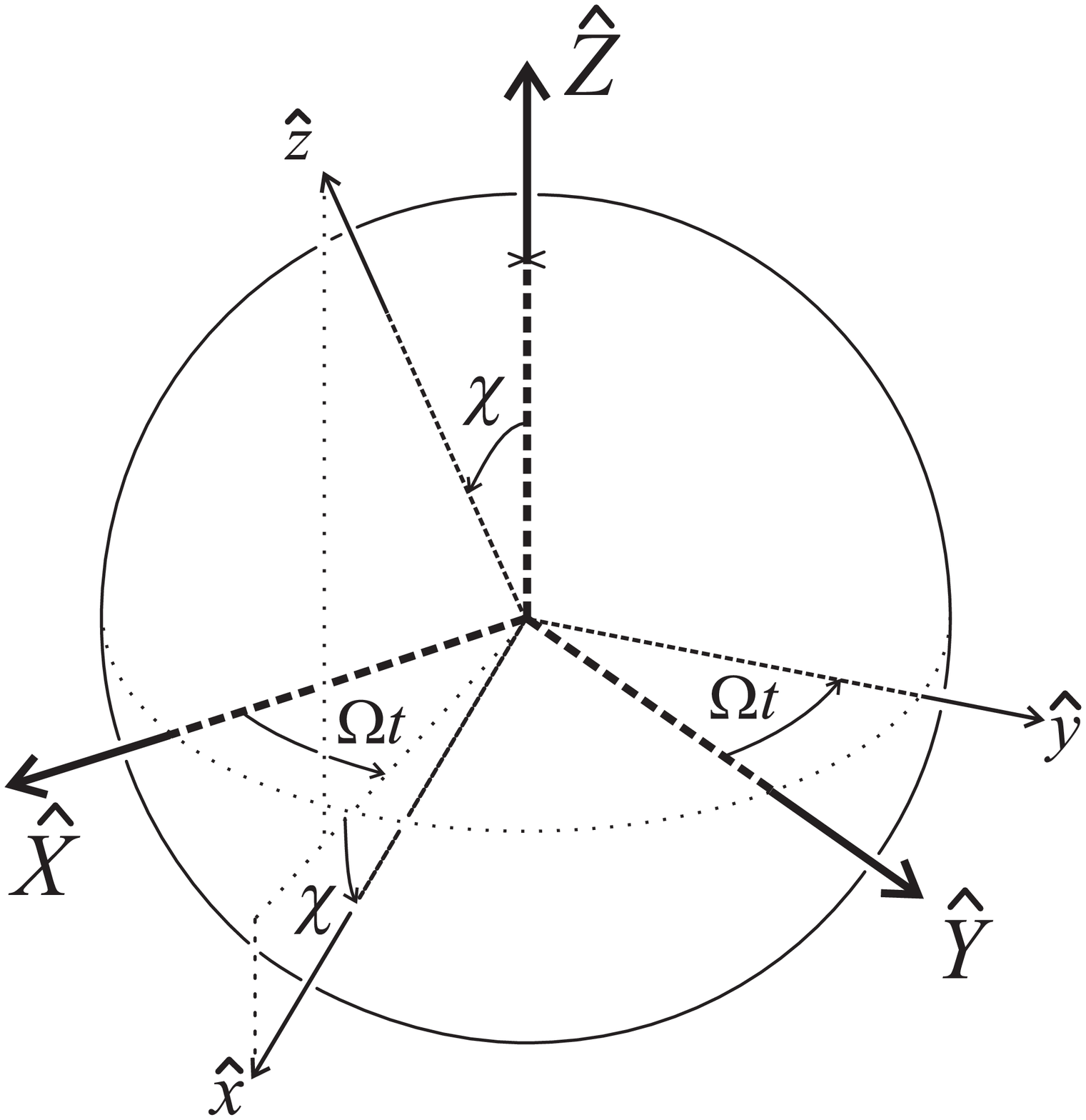}  
    \includegraphics[width=0.40\textwidth]{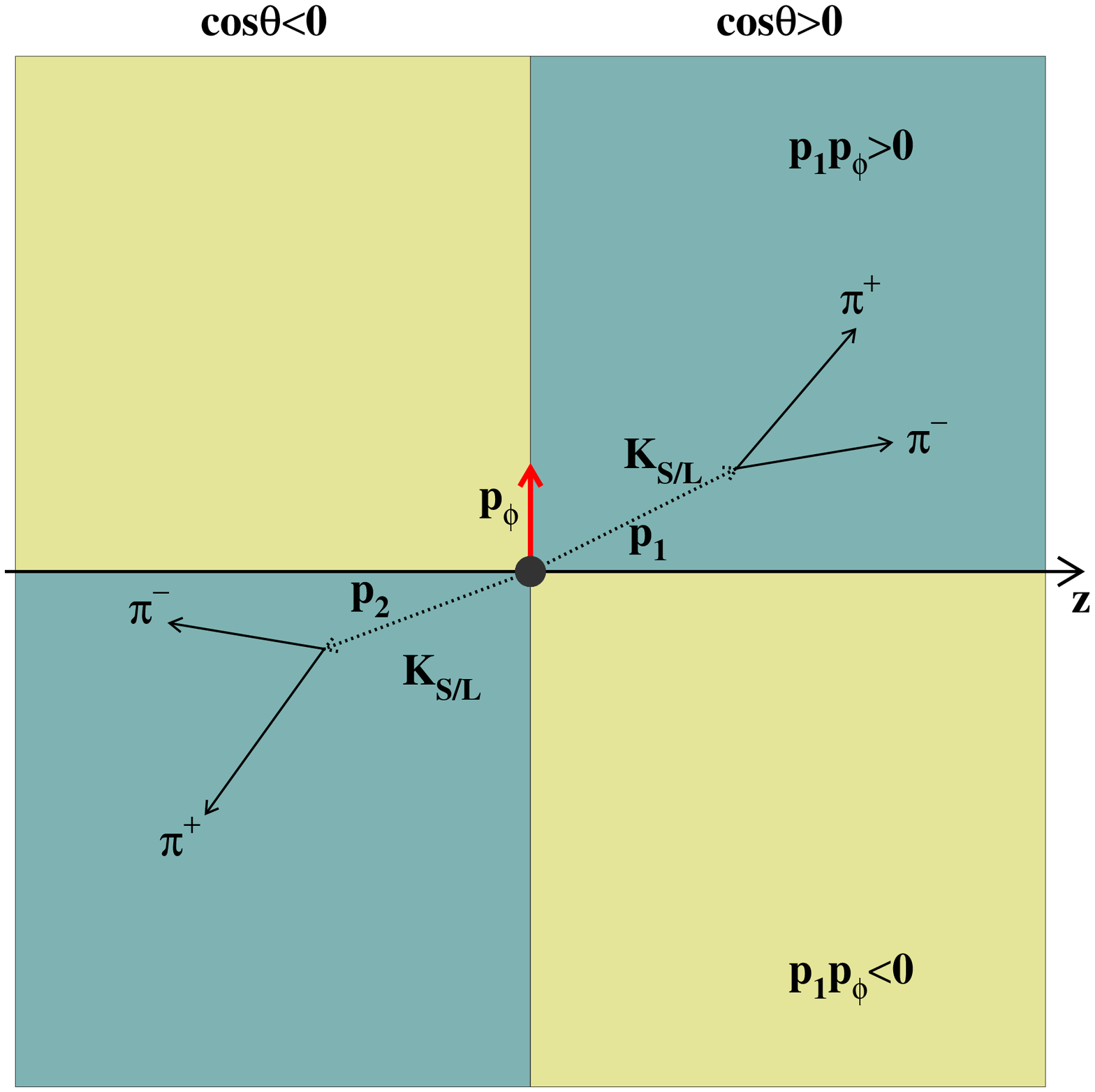}
  \end{center}
  \caption{Left: Basis $(\hat{x},\hat{y},\hat{z})$ for the lab frame, 
    and basis $(\hat{X},\hat{Y},\hat{Z})$ for the fixed stars frame. 
    The laboratory frame precesses around the Earth's rotation axis $\hat{Z}$
    at the sidereal frequency $\Omega$. The angle between $\hat{Z}$ and the
    positron beam direction $\hat{z}$ defined in the laboratory frame of 
    KLOE is $\chi \simeq 113^{\circ}$ (picture taken from \Ref{kost2}). 
    Right: Sketch of the quadrant subdivisions as seen from a top view of the 
    KLOE detector; \pfi\ is lying in the horizontal plane.} 
  \label{frames}
\end{figure}

The physical neutral kaon states are expressed in terms of the flavor 
states as $\ket{\ksl}\propto (1+\epsilon_{S,L})\ket{\ko}\pm 
(1-\epsilon_{S,L})\ket{\ok}~,$
with $\epsilon_{\mathrm{S,L}}=\epsilon_{\mathrm{K}}\pm\deltak$ being the \CP\ 
impurities ($\epsilon_{\mathrm{K}}$ 
and \deltak\ are the usual \T\ and \CPT\ violation parameters).
In SME \CPT\ 
violation manifests to lowest order only in the mixing parameter 
\deltak, (i.e. vanishes at first order in the decay amplitudes), and 
exhibits a dependence on the 4-momentum of the kaon \cite{kost2}:
\begin{equation}
\label{eq:deltak}
\deltak \approx i \sin \phi_\mathrm{SW} e^{i \phi_\mathrm{SW}} \gamma_\mathrm{K} 
(\Delta a_0-\vec{\beta_\mathrm{K}}\cdot \Delta{\vec{a}})/\Delta m \, ,
\end{equation}  
where $\gamma_\mathrm{K}$ and $\vec{\beta_\mathrm{K}}$ are the kaon boost factor 
and velocity in the laboratory frame, 
$\phi_\mathrm{SW}=\arctan (2\Delta m/\Delta\Gamma)$ is the 
so-called {\it superweak} phase, $\Delta m = m_\mathrm{L}-m_\mathrm{S}$, 
$\Delta \Gamma = \Gamma_\mathrm{S}-\Gamma_\mathrm{L}$ are the mass and width 
differences for the neutral 
kaon mass eigenstates, and $\Delta a_{\mu}$ are four \CPT\ and 
Lorentz violating coefficients for the two valence quarks in the 
kaon\footnote{\Damu\ parameters are associated with differences of terms 
in the SME of the form
$-a_{\mu}^{q}\bar{q}\gamma^\mu q$, where $q$ is a quark field of a specific 
flavor \cite{kost2}.}.
The natural choice of the reference frame to observe these parameters is the 
reference frame defined by fixed stars. Following \Ref{kost2} and choosing a 
three-dimensional basis ($\hat{X},\hat{Y},\hat{Z}$) in this frame 
(with $\hat{Z}$ parallel to Earth's rotation axis, $\hat{X}$ expressed in terms
of celestial equatorial coordinates with declination and right ascension 
$0^\circ$, while $\hat{Y}$ with declination $0^\circ$ and right ascension 
$90^\circ$) and a basis $(\hat{x},\hat{y},\hat{z})$ for the laboratory frame 
according to \fig{frames}-left, (where $\hat{z}$ is parallel to the beam axis 
with direction fixed by the positron beam and $\hat{x}$ lies in the plane 
$(\hat{z},\hat{Z})$), the \CPT\ violating parameter $\deltak$ is expressed as:
\begin{eqnarray} 
\deltak(\vec{p},\Ts) &=& 
\frac{i \sin\phi_\mathrm{SW}e^{i\phi_\mathrm{SW}}}{\Delta m} \gamma_\mathrm{K}
\Big[ \Dao \nonumber \\
&&
+ \beta_\mathrm{K} { \Daz}(\cos\vartheta\cos\chi-\sin\vartheta\cos\varphi\sin\chi)
\nonumber \\
&&
-\beta_\mathrm{K} \Dax \sin\vartheta\sin\varphi\sin\Omega \Ts
\nonumber \\
&& 
+ \beta_\mathrm{K} \Delta
 a_X(\cos\vartheta\sin\chi+\sin\vartheta\cos\varphi\cos\chi)\cos\Omega \Ts
\nonumber \\
&&
+ \beta_\mathrm{K} \Day 
 (\cos\vartheta\sin\chi+\sin\vartheta\cos\varphi\cos\chi)\sin\Omega \Ts
\nonumber \\
&&
+\beta_\mathrm{K} \Day\sin\vartheta\sin\varphi\cos\Omega \Ts
\Big]
\label{eq:intro:delta_sid}
\end{eqnarray}
where $\vec{p}$ is the kaon momentum, $\Ts$ is the sidereal time, $\Omega$ 
is the Earth's sidereal frequency, $\cos\chi=\hat{z}\cdot\hat{Z}$; 
$\vartheta$ and $\varphi$ are the conventional polar and azimuthal angles of 
the kaon momentum in the laboratory frame.

The sensitivity to the four \Damu\ parameters can be very different 
for fixed target and collider experiments, showing complementary features 
\cite{kost2,ref:HandbookAD}. At a fixed target experiment the kaon momentum 
direction is fixed, while $|\vec{p}|$ might vary within a certain interval. 
On the contrary, at a \ff\ kaons are emitted in all directions
with the characteristic $p$-wave angular distribution 
$\dd N/\dd\Omega \propto 
\sinp{2}{\vartheta}$, while $|\vec{p}|$ is almost fixed\footnote{At \dafne\ 
  $|\vec{p}|$ is not fixed because of a small \f\ meson momentum 
  \pfi\ in the laboratory frame ($|\pfi|\simeq15$ \MeV).}.
The analysis strategy used to measure the four \Damu\ parameters exploits 
the entanglement of neutral kaons produced at \dafne\ \cite{ref:HandbookAD}. 
In fact, at a \ff, 
neutral kaons 
are produced in pairs in a coherent quantum state with the \f-meson quantum 
numbers $\JPC=1^{-\,-}$:
\begin{eqnarray} 
\ket{i} & = & \frac{1}{\sqrt{2}}(\ket{\ko,\puno}\ket{\ok,\pdue}-
\ket{\ok,\puno}\ket{\ko,\pdue}) = \nonumber \\ 
~ & ~ & \frac{\mathcal{N}}{\sqrt{2}}(\ket{\ks,\puno}\ket{\kl,\pdue}-
\ket{\kl,\puno}\ket{\ks,\pdue}) \label{eq:state}
\end{eqnarray}
where $\mathcal{N}=
{\sqrt{(1+|\epsilon_\mathrm{S}|^2)(1+|\epsilon_\mathrm{L}|^2)}}/
{(1-\epsilon_\mathrm{S}\epsilon_\mathrm{L})} 
\simeq 1$ is a normalization factor and \puno(\pdue) are the kaon momenta.
\par
The observable quantity is the double differential decay rate of the state 
$|i\rangle$ into decay products $f_1$  and $f_2$ at proper times $\tau_1$ and
$\tau_2$, respectively.
After integration on $(\tau_1+\tau_2)$ at fixed time difference 
$\Dt=\tau_1-\tau_2$, the decay intensity can be written as follows 
\cite{ref:HandbookAD}:

\begin{equation}
\label{eq:intro:timeevo} 
  I_{f_1f_2}(\Dt)   =  
 C_{12} ~ e^{-\Gamma |\Dt|} \Big[ 
    |\eta_{1}|^2 e^{\frac{\Delta\Gamma}{2} \Dt}  + 
    \,\,|\eta_{2}|^2 e^{-\frac{\Delta \Gamma}{2}\Dt}
    -2 \Re\Big(\eta_{1}\eta_{2}^*e^{-i\Delta m \Dt} \Big) \Big] 
\end{equation}
with $\Gamma = (\Gamma_\mathrm{S}+\Gamma_\mathrm{L})/2$,
\begin{eqnarray}
\eta_j &\equiv& |\eta_j|e^{i\phi_j} = 
\frac{\braket{f_j}{T|\kl}}{\braket{f_j}{T|\ks}}~,\nonumber \\
C_{12}&=&\frac{ |\mathcal{N}|^2 }{2 (\Gamma_\mathrm{S}+\Gamma_\mathrm{L})} 
|\braket{f_1}{T|\ks}\braket{f_2}{T|\ks}|^2~.\nonumber
\end{eqnarray}  

In particular when $f_1=f_2=\pi^+\pi^-$ ($I_{f_1,f_2}(\Dt)\equiv \Ipm$) 
the corresponding $\eta_j$ parameters can be slightly different for the two 
kaons due to the momentum dependence of the \CPT\ violation effects arising 
from \equ{eq:intro:delta_sid}:
\begin{eqnarray} 
  \eta_1 &  \simeq & \epsilon_K-\deltak(\puno,\Ts) \nonumber\\
  \eta_2 & \simeq & \epsilon_K-\deltak(\pdue,\Ts)~, \label{eq:intro:etapm} 
\end{eqnarray}
with $\pdue=\pfi -\puno$. Higher order contributions to CP violation 
common to the two $\eta_j$ coefficients have been neglected.
Due to the presence of the fully destructive quantum interference
at $\Dt=0$, the  distribution $\Ipm$ is extremely sensitive to any deviation 
from unity of the ratio $\eta_1/\eta_2$ in this interference region\footnote{
  Outside the interference region $|\Dt|\gg\tau_\mathrm{S}$ the distribution 
  \Ipm\ is
  only sensitive to deviations from unity of the ratio $|\eta_1/\eta_2|^2$.
  In this case no \CPT\ violation effect can be observed because $\epsilon_K$
  and $\deltak$ are expected to be $90^\circ$ out of phase 
(i.e. $\Re e(\deltak/\epsilon_K)\sim 0$).} 
(i.e. $\Dt \approx 0$). Therefore a suitable analysis of the decays
\f\to\ks\kl\to\pip\pim\pip\pim\
as a function of sidereal time and kaon momenta can provide a measurement of 
the four parameters \Damu.
\par
It is worth noting that the presence of the small  momentum \pfi\
makes $\gamma_{1}\neq\gamma_{2}$ on an event-by-event basis, which is a 
necessary condition in order to have the \Ipm\ distribution 
sensitive to the \CPT\ violation effects induced by the \Dao\ parameter.
\par
The two kaons are identified by their emission in the forward 
($\cos\vartheta>0$) or backward ($\cos\vartheta<0$) hemispheres, as shown 
in \fig{frames}-right. The data sample is divided into two subsets in which the 
kaons going in the forward direction ($\cos\vartheta >0$) are emitted in a 
quadrant along with ($\puno\cdot\pfi>0$) or opposite to ($\puno\cdot\pfi<0$) 
the \f\ momentum \pfi, thus having a higher (or lower) value of $\gamma_1$ 
than the companion kaons emitted in the backward direction (see 
\fig{frames}-right).
Moreover the data are divided into four bins of sidereal time. Fitting 
simultaneously the corresponding eight \Ipm\ distributions modulation 
effects induced by the \CPT-violating parameter $\deltak$ in 
\equ{eq:intro:delta_sid} as a function of sidereal time and kaon momentum
can be observed.
\par
The observable decay rate distribution as a function of \Dt\ for these bins 
is:
\begin{equation} \label{eq:intro:observable_theo}
  S_{jh}(\Dt) = \int_{\Delta\Ts_j}\!\!\!\!\! \dd\Ts\,\rho(\Ts)
  \int_{\Delta\Omega_h}\!\!\!\!\!\dd\varphi\dd\vartheta\,g(\vartheta) \,\Ipm
\end{equation}
The function $\rho(\Ts)$ is the sidereal time density of recorded events 
and has been derived from an independent data sample. Indices  
\emph{j}=1,4 and \emph{h}=1,2 are for the sidereal time and angular 
bin, respectively. The factor $g(\vartheta)$ takes into account the 
kaon polar angle distribution in the \f\ meson center of mass frame 
($\sinp{3}{\vartheta}$).
\par 
Starting from \equ{eq:intro:observable_theo} the expected distribution is:
\begin{equation} \label{eq:intro:fitfun}
  \tilde{S}_{jh}(\Dt) = (1+f_\mathrm{regen}(\Dt))\int\!\!\dd\Dt'
  \,\varepsilon_\mathrm{tot}(\Dt')\,P_\mathrm{MC}(\Dt'-\Dt)\,S_{j,h}(\Dt')
\end{equation}
where $f_\mathrm{regen}(\Dt)$ is a correction factor that takes into account 
the kaon regeneration on the beam pipe structures, 
$\varepsilon_\mathrm{tot}(\Dt')$ is the total detection efficiency, 
and $P_\mathrm{MC}(\Dt'-\Dt)$ is the smearing matrix 
for the \Dt\ resolution, both evaluated with 
the Monte Carlo (MC) simulation. The correction induced by kaon regeneration 
has been evaluated according to \Ref{Ambrosino:2006vr}.


\section{The KLOE experiment}\label{sec:kloe}

The KLOE experiment operates at \dafne, the Frascati \ff.\\
\dafne\ is an \ep\el\ collider working at a center of mass energy of 
\ab1020 \MeV, the mass of the \f-meson. 
Positron and electron beams collide at an angle of $\pi$-25 mrad, 
producing \f\ mesons with non-zero momentum in the horizontal plane, 
$|\pfi|\sim15\,\MeV$ in the KLOE reference frame.
\par
The beam pipe at the interaction region of \dafne\ has a spherical shape, 
with a radius of 10 cm, and is made of a aluminum-beryllium alloy 500 
\mum\ thick.
A thin beryllium cylinder, 50 \mum\ thick and 4.4 cm radius, is inserted to
ensure electrical continuity.  
\par
The KLOE detector consists of a large cylindrical drift chamber (DC) 
surrounded by a lead-scintillating fiber electromagnetic calorimeter (EMC). 
A super-conducting coil around the EMC provides a 0.52 T axial field. 
\par
The DC \cite{Adinolfi:2002uk} is 4 m in diameter and 3.3 m long and has 
12,582 all-stereo cells properly arranged in 58 layers to ensure homogeneous 
detector response. Time and amplitude of signals from cells are read-out 
to measure hit positions and energy loss 
\cite{Balla:2006cq}. The chamber structure is made of carbon-fiber epoxy 
composite 
and the gas mixture used is 90\% helium, 10\% isobutane. These features 
maximize transparency to photons and reduce \kl\to\ks\ regeneration and multiple
scattering. The position resolutions for single hits are $\sigma_{r,\phi}$\ab 150
\mum\ and $\sigma_z$\ab 2 \mm\ in the transverse and longitudinal plane, 
respectively and are almost homogeneus in the active volume. The momentum 
resolution is $\sigma(p_{\perp})/p_{\perp}\ab 0.4\%$ for polar angles in the range 
$40^\circ-130^\circ$. 
\par
The EMC \cite{Adinolfi:2002jk} is divided into a barrel and two end-caps, and 
covers 98\% of the solid angle. Signal amplitude and time of the 
modules are read-out at both ends by photo-multipliers for a total of 2440 
cells arranged in five layers in depth. Cells close in time and space are 
grouped 
into calorimeter clusters. The cluster energy $E_\mathrm{clu}$ is the sum of its 
cell energies. 
The cluster time $T_\mathrm{clu}$ and position $\vec{R}_\mathrm{clu}$ 
are energy-weighted 
averages. Energy and time resolutions are $\sigma_E/E_\mathrm{clu} = 
5.7\%/\sqrt{E_\mathrm{clu}(\GeV)}$ and  
$\sigma_{T_\mathrm{clu}} = 57{\mathrm{ps}}/\sqrt{E_\mathrm{clu}(\GeV)} \oplus 100\ 
{\rm \mathrm{ps}}$, 
respectively. 
\par
During data taking \dafne\ beam conditions and detector calibrations are 
constantly monitored in order to guarantee the highest quality of the collected 
data. 


\section{Data analysis} \label{sec:dataana}

A integrated luminosity of 1.7 \invfb\ have been used in this analysis,
corresponding to the KLOE data-set acquired during years 2004-2005. 
Two different MC samples have been used, with equivalent 
integrated luminosity of 3.4 \invfb\ and 17 \invfb, respectively. The former, 
containing all the \f-meson decay channels, have been used for analysis 
optimization, while the latter, containing signal events only, have been 
used for efficiency and \Dt\ resolution determination.
\par
The data reduction starts with the topological identification of the candidate
signal events: two decay vertices with only two tracks each. For each vertex 
the same kinematic criteria are used for sample selection:
\begin{itemize}
\item $|m_\mathrm{trk}-m_\mathrm{K}|<5$ \MeV \\
  where $m_\mathrm{trk}$ is the invariant mass of the kaon reconstructed 
  from the tracks assuming charged pion mass hypothesis: 
  $\vec{p}_{1,2}=\vec{p}_{\pip}+\vec{p}_{\pim}$ and 
  $E_{1,2}=E_{\pip}+E_{\pim}$. Energy of the pions are defined as:
  $E_{\uppi}=$ $\sqrt{m_\uppi^{2}+|\vec{p}_{\uppi}|^{2}}$;
\item $\sqrt{E_\mathrm{miss}^2+|\vec{p}_\mathrm{miss}|^2}< 10$ \MeV \\
  where $\vec{p}_\mathrm{miss} = (\pfi-\vec{p}_{2,1})-\vec{p}_{1,2}$ and 
  $E_\mathrm{miss}=$$(E_\f-E_{2,1})-E_{1,2}$;
\item $-50~\MeV^2 < m^2_\mathrm{miss} < 10~\MeV^2 $ \\
  where $m^2_\mathrm{miss}= E_\mathrm{miss}^2-|\vec{p}_\mathrm{miss}|^2$ 
\item $|p^*_{1,2}-p^*_0|<10~\MeV$\\
  where $p^*_{1,2}$ is the momentum of the kaon, as derived from tracks, in 
  the \f-meson reference frame, while $p^*_0 = \sqrt{s/4-m^2_\mathrm{K}}$ 
  and $\sqrt{s}$ is the center-of-mass energy as measured run-by-run 
  with large angle Bhabha scattering events.
\end{itemize}
\par
In the analysis the two kaons are ordered according to the value of the 
longitudinal momentum, $p_z$. To avoid the region where this order is diluted
by resolution effects, events with $p_{1,z} < 2\,\MeV$ are rejected.
\par
In the standard data reconstruction \cite{Ambrosino:2004qx} the position 
of the kaon decay vertex is determined using a purely geometrical approach 
(\emph{i.e.}, without kinematical information) and each vertex is fit 
independently of the rest of the event. In order to improve the accuracy of 
decay time measurements a dedicated fit procedure has been developed for this
analysis. The free parameters of the fit are the \f-meson 
decay point coordinates ($\vec{V}_\f$) and the decay  length for the two kaons 
($\lambda_{1,2}$). The two vertices are assumed to be of the form 
$\vec{V}_{1,2}=\vec{V}_\f+\lambda_{1,2}\hat{n}_{1,2}$, where 
$\hat{n}_{1,2}=\vec{p}_{1,2}/|\vec{p}_{1,2}|$ is the flight direction of the kaon.
A global Likelihood function is built in order to take into 
account both vertices at the same time:
$$
-2\log{\mathcal{L}} = 
\sum_{i=1,2}\log{P_{i}(\vec{V}_i(\mathrm{rec})-\vec{V}_i(\mathrm{fit}))} +
\log{P_{\f}(\vec{V}_\f(\mathrm{rec})-\vec{V}_\f(\mathrm{fit}))}~,
$$
where $P_j(P_\f)$ are the probability density function for displacements with 
respect to the reconstructed positions as derived from MC. 
Events with \\$-2\log{\mathcal{L}}>30$ or with pulls on decay length 
$|\lambda_{j}(rec)-\lambda_{j}(fit)|/\sigma_j(fit)>10$ are rejected. 
\par
\begin{figure}[htb]
  \begin{center}
    \includegraphics[width=0.48\textwidth]{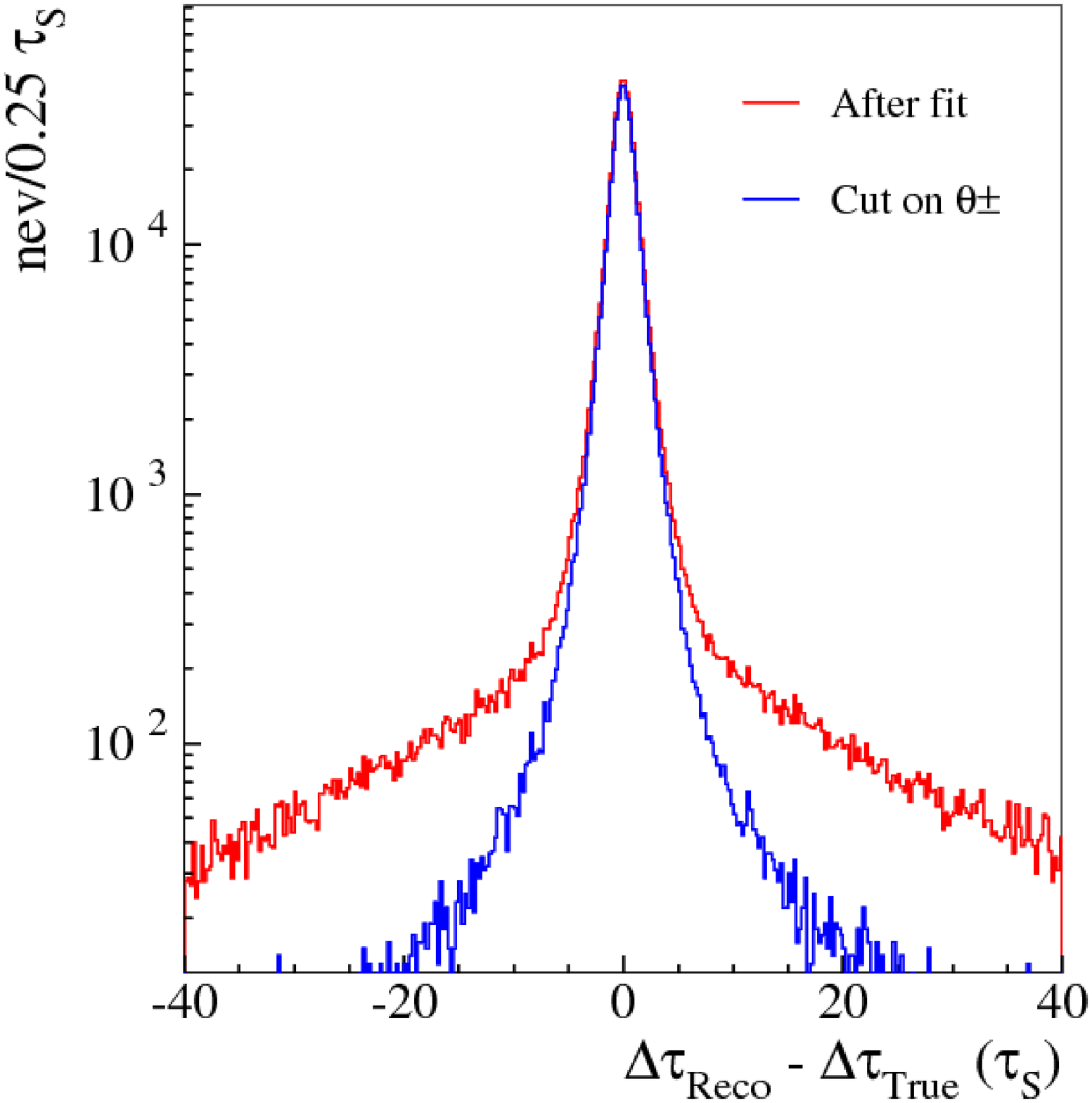}
    \includegraphics[width=0.48\textwidth]{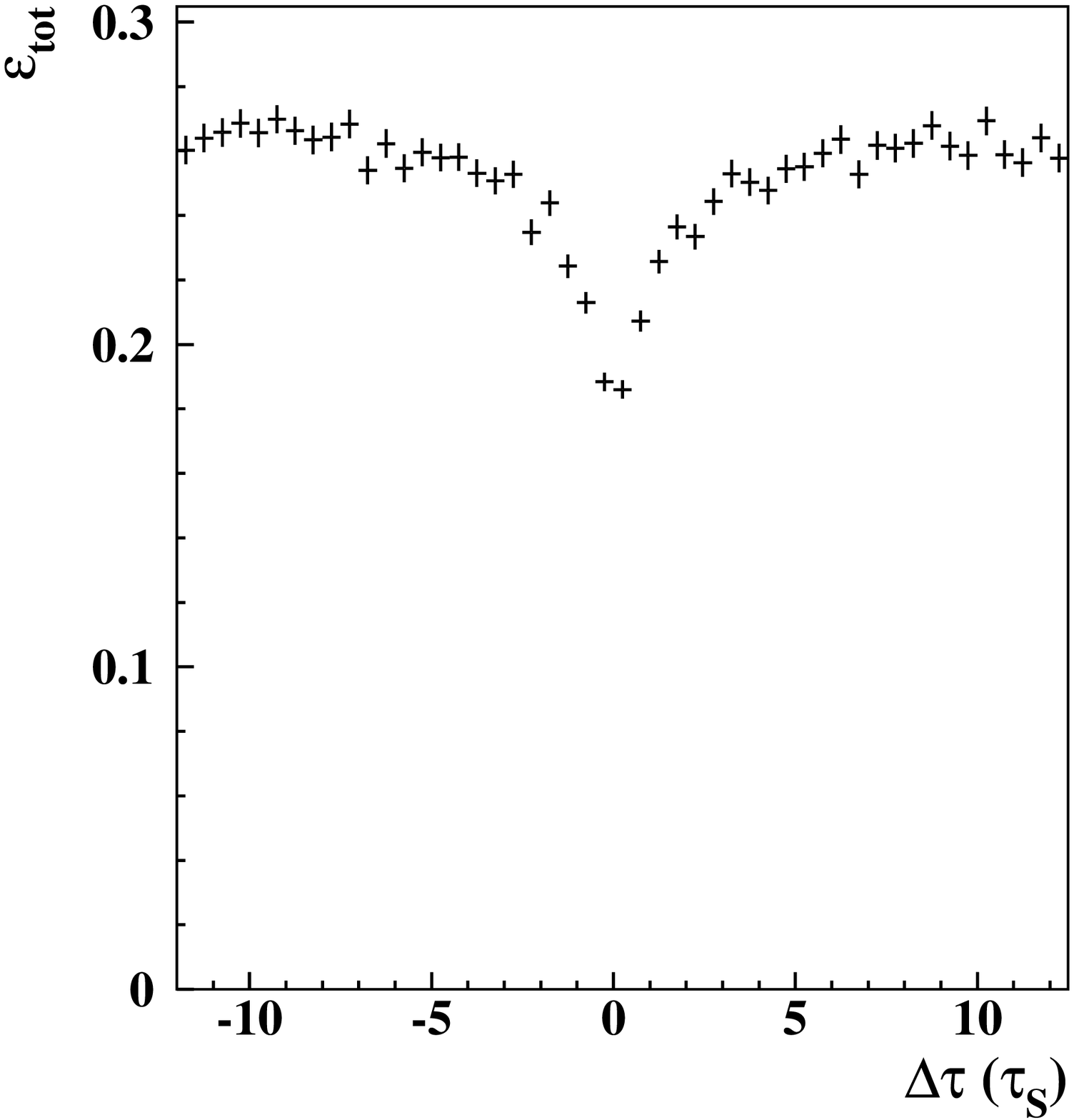}
  \end{center}
  \caption{Left: resolution on \Dt\ variable evaluated from MC simulation
    as the difference between true values and data-like reconstructed values. 
    The effect of the cut on the opening angle between tracks is shown. 
    Right: total detection efficiency as a function of true \Dt\ as
    observed from MC simulation.The dip around $\Dt \sim 0$ is discussed 
    in the text.}
  \label{fig:dataana:mceffi}
\end{figure}

The \Dt\ resolution is strongly correlated, as expected, with the opening angle 
of the pion tracks ($\vartheta_\pm$), and deteriorates at large values of 
$\vartheta_\pm$. A cut to eliminate large opening angles values has been 
applied and events with $\cos\vartheta_\pm <-0.975$ are rejected. The 
result is shown in \fig{fig:dataana:mceffi}-left.
The core width of \Dt\ resolution after all selection is 
$\sim\!0.6\,\tau_\mathrm{S}$ as estimated by fitting the MC simulation 
distribution with the superposition of three 
Gaussian functions. The agreement with experimental data has been verified by 
comparing the kinematic fit pull distributions on decay length between 
experimental data and MC simulation.

The main background source is the kaon regeneration on the spherical beam pipe.
In order to reduce this contribution the present analysis is restricted only 
to events with both \ks\ and \kl\ decaying inside the beam pipe. This 
geometrical requirement translates in maximal absolute value for \Dt\ of 
$12\tau_\mathrm{S}$. In this case kaon regeneration occurs only in the thin 
beryllium cylinder.
\par
The residual background contamination in the \Dt\ range 
$[-12;12] \tau_\mathrm{S}$, as estimated from MC simulation, is 2\% from 
regeneration on the beryllium cylindrical foil and 0.5\% 
$\ep\el\to\pip\pim\pip\pim (4\uppi)$ direct production as estimated from a 
dedicated data analysis. The latter is concentrated around $\Dt\simeq0$ while 
the former will be treated as a multiplicative correction applied to the 
fitting function as described in \equ{eq:intro:fitfun}. 
The $4\uppi$ contribution is estimated to be $54\pm5$ events, according to 
\Ref{Ambrosino:2006vr}.
\par
For the purpose of our measurement, precise knowledge of the total detection 
efficiency value ($\varepsilon_\mathrm{tot}$) is not required; it is 
however important to keep under control its dependence on \Dt, shown in 
\fig{fig:dataana:mceffi}-right. The mean value is 25\% for \Dt\ 
ranging in $[-12;12] \tau_\mathrm{S}$ with a consistent drop for $\Dt \sim 0$. 
This is due to two concurrent effects: increasing extrapolation length 
for tracks coming from decays very close to the beam interaction point and 
wrong association of tracks to the kaon decay vertices\footnote{When the two 
vertices are so close in space ($\Dt \ab 1\tau_\mathrm{S} \Rightarrow 
\Delta \lambda_V \sim 6 \mm$) it is possible to wrongly associate tracks to 
vertex by exchanging two of them. These vertices are then rejected because of 
the kinematical requirements.}. 
The efficiency $\varepsilon_{tot}$ can be divided into three main components, 
related to the trigger, to the reconstruction procedure and to the effect of \
the selection criteria:
$$
\varepsilon_\mathrm{tot} = \varepsilon_\mathrm{trig}\varepsilon_\mathrm{reco}
\varepsilon_\mathrm{cuts}
$$
The last contribution is derived from MC simulation while for trigger and 
reconstruction efficiencies, corrections obtained from real data are used. 
To this aim a high purity, independent control sample of \ks\to\pip\pim\ 
and \kl\to\pion\muon\neutrino\ has been selected with particles in the same 
momentum range as for the signal.
  
The purity of the control sample is 95\%, the background being  
dominated by \ketre\ decay. The efficiency correction has been evaluated as 
the ratio between data and MC distributions of $I_{{K_\mu3}\pm}(\Dt)$. The decay 
time ordering criterion is the same used in the main analysis.
\par
\begin{figure}[htb]
  \begin{center}
    \includegraphics[width=0.99\textwidth]{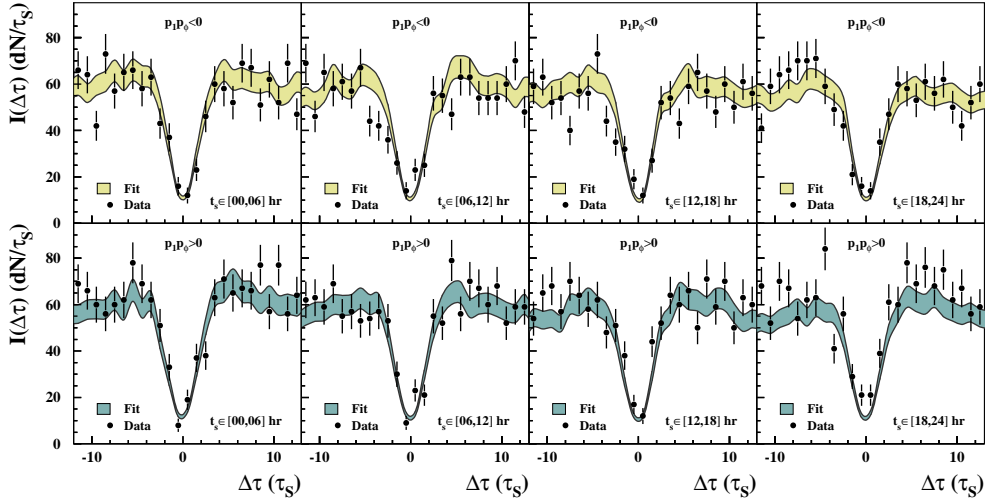}
  \end{center}
  \caption{Fit results: the top and bottom plots refer to the two 
    angular selections. Black points are for experimental data while 
    colored bands are the fit output. 
    The error bars on experimental data are purely statistical, while the band 
    represents the contribution to the uncertainty due to MC 
    simulation statistics and efficiency correction.} 
  \label{fig:dataana:fit}
\end{figure}

Experimental data distributions have been analyzed for different intervals of
sidereal time and kaon emission angle using the function in 
\equ{eq:intro:fitfun}. The \Dt\ range is $\Dt \in [-12:12]\tau_\mathrm{S}$ 
with $1\,\tau_\mathrm{S}$ bin width, while sidereal time has 4 bins, 6 hours 
each, and two angular  bins have been used: $\puno\cdot\pfi \gtrless 0$ 
resulting in a total of 192 experimental points simultaneously fit. The fit 
$\chi^2/N_\mathrm{DoF}$ is $211.7/187$ corresponding to a probability of 10\%,
numerical values of the fit are reported in \tab{tab:dataana:results} and the 
experimental data distributions are shown in \fig{fig:dataana:fit}.

\begin{table}[!htb]
  \caption{Fit results. Errors include all source of the statistical 
    fluctuations. The fit $\chi^2/N_\mathrm{DoF}$ is $211.7/187$ corresponding 
    to a probability of 10\%. The table contains the correlation coefficient 
    matrix between the parameters.}
  \label{tab:dataana:results}
  \newcommand{\m}{\hphantom{-}}
  \renewcommand{\tabcolsep}{0.6pc}  
  \renewcommand{\arraystretch}{1.1} 
  \begin{center}
    \begin{tabular}{ l | c c c c } 
      Fit output ($10^{-18} \,\, \GeV$ units) & 
      \multicolumn{4}{c}{Correlation matrix} \\
      \hline                                                                    
      \Dao =  $-6.0 \pm 7.7 $  &   1.000 &  0.304 & -0.187 &  0.483 \\
      \hline
      \Dax =  $\m0.9 \pm 1.5$  &   0.304 &  1.000 & -0.045 &  0.069 \\ 
      \hline
      \Day =  $-2.0 \pm 1.5$  &  -0.187 & -0.045 &  1.000 & -0.104 \\
      \hline
      \Daz =  $\m3.1 \pm 1.7$  &   0.483 &  0.069 & -0.104 &  1.000 \\
      \hline
    \end{tabular}
  \end{center}
\end{table}

Systematic effects are listed in \tab{tab:dataana:sys}. The full analysis 
chain has been repeated several times varying all the cuts values according 
to the MC resolution ($\sigma_\mathrm{MC}$) of the selection variables. 
Positive and negative variations ($\pm\sigma_\mathrm{MC}$) have been considered.
The results of the fits are stable and the RMS of the distribution of obtained
results is taken as systematic uncertainty. 
\par
The accuracy on the determination of the \Dao\ parameter strongly depends on 
the stability of the observed \Ipm\ distribution for $|\Dt|>5\tau_\mathrm{S}$. 
For this reason the fit range has been varied of $\pm 1\tau_\mathrm{S}$ around 
the reference value enlarging or reducing the \Dt\ range. Correspondingly 
the event yield varies of $\sim 10\%$ and the RMS of the fit results has been 
chosen as the related systematic error. The largest effect, as expected, has 
been observed on \Dao. 
\par
The $4\uppi$ background events is concentrated in the two bins around 
$\Dt\sim0$, resulting in a bin-by-bin contribution of $\sim3$ events, while 
the amount of observed data events in the same bins is around $\sim10$ events. 
Being the amount of subtracted events similar to the statistical uncertainty 
of the observed data the systematic effect of the subtraction has been obtained
as the difference between results with and without subtraction.
\par
The effect of the orientation of the KLOE reference frame with respect to the 
terrestrial coordinate system has been taken into account in the fit function.
The measurement of the relative alignment between our reference frame with 
respect to the geographical frame has been performed with a compass and the 
effect of displacement between magnetic and true North pole has been taken into 
account and corrected for. The angle between the laboratory $\hat{z}$ axis and 
the North direction is $(130\pm2)^\circ$ on the local tangent plane. 
The fit function has been evaluated with angle variations up to $10^\circ$ 
showing stable results. The RMS of the results has been taken as 
systematic uncertainty.
\par
The kaon regeneration correction in \equ{eq:intro:fitfun} has been varied 
according to its relative uncertainty ($\sim 35\%$). The corresponding results 
fluctuation is negligible with respect to the other systematic uncertainty. 
\par
The effect of mismatch in \Dt\ resolution between experimental data and MC 
simulation has been found negligible inside the maximum allowable error of 
5\% on \Dt\ resolution width. 
\par
The final systematics uncertainty is the sum in quadrature of all the 
discussed effects and is reported in the last column of \tab{tab:dataana:sys}. 
In all cases the total systematic uncertainty ranges between 30\% and 40\% 
of the corresponding statistical uncertainty.
\begin{table}[!htb]
  \caption{Summary of the systematic uncertainty in $10^{-18}\,\GeV$ units.}
  \label{tab:dataana:sys}
  \newcommand{\m}{\hphantom{-}}
  \renewcommand{\tabcolsep}{0.6pc}  
  \renewcommand{\arraystretch}{1.1} 
  \begin{center}
    \begin{tabular}{r|c|c|c|c|l  } 
      ~  & Analysis cut & \Dt\ Range & $4\pi$ subtr. & Ref. frame & Total \\
      \hline    
      $\Dao$ &  1.1 & 2.4 & 1.3 & 1.0 & 3.1 \\
      \hline    
      $\Dax$ &  0.3 & 0.3 & 0.4 & 0.2 & 0.6\\
      \hline     
      $\Day$ &  0.2 & 0.3 & 0.2 & 0.2 & 0.5 \\
      \hline    
      $\Daz$ & 0.2 & 0.2 & 0.4 & 0.4 & 0.6\\
      \hline 
    \end{tabular}
  \end{center}
\end{table}
   
\section{Results and conclusions} \label{sec:conclusions}
The results for the \Damu\ parameters are:\\
\begin{center}
$\Dao = (-6.0 \pm 7.7_{stat} \pm 3.1_{syst})\times 10^{-18} \,\, \GeV$,  \\
$\Dax = (\,\,\,0.9 \pm 1.5_{stat} \pm 0.6_{syst})\times 10^{-18} \,\, \GeV$,  \\
$\Day = (-2.0 \pm 1.5_{stat} \pm 0.5_{syst})\times 10^{-18} \,\,\GeV$,  \\
$\Daz = (\,\,\,3.1 \pm 1.7_{stat} \pm 0.5_{syst})\times 10^{-18} \,\,\GeV$.\
\end{center}
\par
The systematic uncertainty is smaller than the corresponding statistical 
error implying that the main limitation of the present measurement is the 
available statistics. To this respect the continuation of the KLOE physics 
program in the framework of the KLOE-2 experiment \cite{AmelinoCamelia:2010me}
at upgraded \dafne\ machine \cite{Zobov:2010zza} is important to further
improve the results.  
\par
The present result is the first complete measurement of all \Damu\ 
parameters in the kaon sector\footnote{Preliminary results from KTeV 
collaboration in this sector can be found in the \Ref{Nguyen:2001tg}.}. 
The order of magnitude of the results is approaching the interesting region 
defined by $m_{\mathrm{K}}^2/\mathcal{M}_\mathrm{Plank} \sim 2\times10^{-20}\,\GeV$ 
in the assumption that the SME phenomenology of \CPT\ and Lorentz invariance 
breaking is related to some underlying quantum gravity effects. 
\par
These results can be compared to similar ones for the corresponding \Damu\ 
parameters for the B and D neutral meson systems 
\cite{Aubert:2007bp,Link:2002fg} which have a precision of \order{-13}, \GeV.

\section*{Acknowledgments} 
We warmly thank our former KLOE colleagues for the access
to the data collected during the KLOE data taking campaign. We
thank the \dafne\ team for their efforts in maintaining low background
running conditions and their collaboration during all data
taking. We want to thank our technical staff: G.F. Fortugno and
F. Sborzacchi for their dedication in ensuring efficient operation
of the KLOE computing facilities; M. Anelli for his continuous attention
to the gas system and detector safety; A. Balla, M. Gatta,
G. Corradi and G. Papalino for electronics maintenance; M. Santoni,
G. Paoluzzi and R. Rosellini for general detector support;
C. Piscitelli for his help during major maintenance periods.
We acknowledge the support of the European Community –
Research Infrastructure Integrating Activity ‘Study of Strongly Interacting
Matter’ (acronym HadronPhysics2, Grant Agreement No.
227431) under the Seventh Framework Programme of EU. This
work was supported also in part by the EU Integrated Infrastructure
Initiative Hadron Physics Project under contract number
RII3-CT-2004-506078; by the European Commission under the 7th
Framework Programme through the ‘Research Infrastructures’ action
of the ‘Capacities’ Programme, Call: FP7-INFRASTRUCTURES-
2008-1, Grant Agreement No. 283286; by the Polish National Science
Centre through the Grants No. 0469/B/H03/2009/37, 0309/B/H03/2011/40, 
DEC-2011/03/N/ST2/02641, 2011/01/D/ST2/00748,
2011/03/N/ST2/02652,2011/03/N/ST2/02641, 2013/08/M/ST2/00323 
 and by the Foundation for Polish Science through 
the MPD programme and the project HOMING PLUS BIS/2011-4/3.

\bibliographystyle{elsarticle-num}

\begin{thebibliography}{99}

%
\bibitem{cpttheo}
  G. Lueders, Ann. Phys. (NY) {\bf 2} (1957) 1, reprinted in Ann. Phys.
  (NY) {\bf 281} (2000) 1004; 
  W.~Pauli, {\it Exclusion principle, Lorentz group and reflexion of space-time
    and charge} in {\it Niels Bohr and the development of physics}, edited by 
  W.~Pauli, Pergamon, London, 1955, p.30; 
  J.~S.~Bell, Proc. R. Soc. London A {\bf 231} (1955) 479;
  R.~Jost, Helv. Phys. Acta {\bf 30} (1957) 409.

\bibitem{Greenberg:2002uu}
  O.~W.~Greenberg,
  Phys.\ Rev.\ Lett.\  {\bf 89} (2002) 231602.

\bibitem{gravity}
  V.~A.~Kosteleck\'y and S.~Samuel, Phys.\ Rev.\ D {\bf 39}, 683 (1989);\\
  V.~A.~Kosteleck\'y and R.~Potting, Nucl.\ Phys. B {\bf 359}, 545 (1991).

\bibitem{Colladay:1996iz}
  D.~Colladay and V.~A.~Kosteleck\'y,Phys.\ Rev.\ D {\bf 55} (1997) 6760;\\
  D.~Colladay and V.~A.~Kosteleck\'y,Phys.\ Rev.\ D {\bf 58} (1998) 116002;\\
  V.~A.~Kosteleck\'y and R.~Potting, Phys.\ Rev.\ D {\bf 51} (1995) 3923. 

\bibitem{Kostelecky:2008ts}
  V.~A.~Kosteleck\'y and N.~Russell,
  Rev.\ Mod.\ Phys.\  {\bf 83} (2011) 11,
  (Updates on [arXiv:0801.0287 [hep-ph]]).

\bibitem{kost2} 
  V.~A.~Kosteleck\'y, Phys.\ Rev.\ Lett.\ {\bf 80}, 1818 (1998);\\
  V.~A.~Kosteleck\'y, Phys.\ Rev.\ D {\bf 61}, 016002 (1999);\\
  V.~A.~Kosteleck\'y, Phys.\ Rev.\ D {\bf 64}, 076001 (2001).


\bibitem{ref:HandbookAD} 
  A. Di Domenico (Editor), 
  Handbook on Neutral Kaon Interferometry at a $\phi$-factory, 
  Frascati Phys. Ser. {\bf 43} (2007).

\bibitem{Ambrosino:2006vr}
  F.~Ambrosino {\it et al.}  [KLOE Coll.],
  Phys.\ Lett.\ B {\bf 642} (2006) 315.

\bibitem{Adinolfi:2002uk} 
  M. Adinolfi \etal, 
  Nucl.\ Inst.\ Meth.\ A {\bf 488} (2002) 51.

\bibitem{Balla:2006cq}
  A.~Balla, M.~Beretta, P.~Branchini, P.~Ciambrone, G.~Corradi, E.~De Lucia, 
  P.~De Simone and G.~Felici {\it et al.},
  Nucl.\ Instrum.\ Meth.\ A {\bf 562} (2006) 141.

\bibitem{Adinolfi:2002jk} 
  M. Adinolfi \etal,
  Nucl.\ Instrum.\ Meth.\ A {\bf 482} (2002) 363.

\bibitem{AmelinoCamelia:2010me}
  G.~Amelino-Camelia \etal,
  Eur.\ Phys.\ J.\ C {\bf 68} (2010) 619.

\bibitem{Ambrosino:2004qx}
  F.~Ambrosino, \etal, [KLOE Coll.]
  Nucl.\ Instrum.\ Meth.\ A {\bf 534} (2004) 403.

\bibitem{Zobov:2010zza}
  M.~Zobov, D.~Alesini, M.~E.~Biagini, C.~Biscari, A.~Bocci, M.~Boscolo, F.~Bossi and B.~Buonomo {\it et al.},
  Phys.\ Rev.\ Lett.\  {\bf 104} (2010) 174801.

\bibitem{Nguyen:2001tg}
  H.~Nguyen,
  ``CPT results from KTeV,''
  unpublished,
  [hep-ex/0112046].

\bibitem{Aubert:2007bp}
  B.~Aubert {\it et al.}  [BaBar Coll.],
  Phys.\ Rev.\ Lett.\  {\bf 100} (2008) 131802.

\bibitem{Link:2002fg}
  J.~M.~Link, \etal [FOCUS Coll.],
  Phys.\ Lett.\ B {\bf 556} (2003) 7.

\end{thebibliography}

\end{document}
